\documentclass[twocolumn,aps,prl,showpacs]{revtex4}
\usepackage[dvips]{graphicx}
\begin{document}
\title{Exact time autocorrelation function of the $N$-spin classical Heisenberg
equivalent neighbor model}
\author{Richard A. Klemm}
\email{rklemm@mpipks-dresden.mpg.de}
\author{Marco Ameduri}
\email{marco@mpipks-dresden.mpg.de}
\affiliation{Max-Planck-Institut f{\"u}r Physik komplexer Systeme,
N{\"o}thnitzer
 Stra{\ss}e 38, D-01187 Dresden, Germany}
\date{\today}
\begin{abstract}
We reduce the  
autocorrelation function ${\cal C}_{11}(t)$ of the equivalent
neighbor
model of $N$ classical spins exhibiting Heisenberg dynamics and exchange coupling
$J$ to quadrature.   As the temperature $T\rightarrow\infty$, ${\cal C}_{11}(t)\propto
t^{-N}$ for $Jt>>1$.
 At low $T$, the antiferromagnetic ${\cal C}_{11}(t)$
is a simple function of $(JT)^{1/2}t$, exhibiting strong frustration, but the
ferromagnetic ${\cal C}_{11}(t)$ oscillates
in a single mode the  frequency of which approaches $NJ$ as $T\rightarrow0$. We conjecture
that as $T\rightarrow\infty$, the  near-neighbor correlation functions of
$N$-spin classical Heisenberg rings are
simply obtained from these  results.
\end{abstract}
\pacs{05.20.-y, 75.10.Hk, 75.75.+a, 05.45.-a}
\vskip0pt\vskip0pt
\maketitle

Recently, there has been a considerable interest  in the physics of magnetic
molecules. \cite{Mn12}  These consist of small clusters of magnetic ions imbedded within a
non-magnetic ligand group, which may crystalize into large, well-ordered
single crystals of sufficient quality for neutron scattering
studies.  Each type of magnetic molecule is characterized not only by the
number $N$ of spins in the molecule, but also by the spatial configuration of
the spins.  Many examples of rings exist, \cite{Fe6,Fe10} but there are also  examples of
denser clusters. \cite{Mn12,Fe8}  Usually, the spins interact mainly via the Heisenberg exchange
interaction.
Although interest in the dynamics of Heisenberg spin rings has been strong for
many years, most of the work involved numerical simulation of the two-spin
correlation functions.  Recently, however, exact results for the Heisenberg
dimer, isosceles and equilateral triangle,  and four-spin ring
were presented, \cite{Luban,KL,AK} 
but larger rings can not be solved using this technique. 

The equivalent neighbor model is the simplest model for the dynamics of a
nanomagnet,  first proposed by
Kittel and Shore.\cite{SK} In this model, every spin
interacts equally with all of the others.   
Although the Heisenberg dynamics of the $N\rightarrow\infty$ limit at infinite temperature $T$ were
obtained previously, \cite{LM}   they
 have not yet been solved for  $5\le N<\infty$.   Although this 
model is 
oversimplified for direct comparision with experiment when $N\ge5$, its exact solution  could serve as a benchmark for future theoretical and experimental studies. 

In this letter, we present an exact single integral representation of the
autocorrelation function ${\cal C}_{11}(t)$ for the equivalent neighbor
model with arbitrary $N$ and $T$,  both for ferromagnetic (FM) and
antiferromagnetic (AFM) exchange couplings. As $T\rightarrow\infty$, 
an accurate asymptotic $1/N$ expansion is found.  As $T\rightarrow0$, the AFM ${\cal
C}_{11}(t)$ 
reduces exactly to a simple function of $(JT)^{1/2}t$ for arbitrary $N$.  For
the FM case at low $T$, ${\cal
C}_{11}(t)$  oscillates in a single mode with a frequency that
approaches $NJ$ as $T\rightarrow0$, and a shape that fits a scaling relation.  From additional studies of
three- and four-spin rings with various amounts of bridging spins that
interact only with the ring spins, we conjecture that 
 the long-time $T\rightarrow\infty$ asymptotic limit values of the
near-neighbor correlation functions for equivalent neighbor coupling and
near-neighbor coupling in classical Heisenberg $N$-spin rings are identical.  
  
  The 
Hamiltonian of the $N$-spin classical Heisenberg equivalent neighbor model is $H=-Js^2/2$, where the total spin ${\bf s}=\sum_{i=1}^N{\bf
s}_i(t)$ is a constant.  The dynamics are given by $d{\bf s}_i(t)/dt=J{\bf
s}_i\times{\bf s}$, where $s_i=|{\bf s}_i|=1$.  ${\bf s}_i(t)$ may be written as
${\bf s}_1(t)=s_{1,||}\hat{\bf s}+s_{1,\perp}[\hat{\bf x}\cos(st^*)-\hat{\bf y}\sin(st^*)],$
where $t^*=Jt$, $\hat{\bf s}=\hat{\bf x}\times\hat{\bf y}$, $s_{1,||}(s,x)=(s^2+1-x^2)/(2s)$,
 $x=|{\bf s}-{\bf s}_1|$, and $s_{1,||}^2+s_{1,\perp}^2=1$. Letting  $\alpha=J/(2k_BT)$, where  $k_B$ is Boltzmann's constant, the
autocorrelation function ${\cal C}_{11}(t)=\langle{\bf s}_1(t)\cdot{\bf
s}_1(0)\rangle$ may be written
\begin{eqnarray}
{\cal C}_{11}(t)&=&{1\over{Z_N}}\int_0^{N-1}{\cal
D}_{N-1}(x)dx\int_{|x-1|}^{x+1}se^{\alpha s^2}ds\times\nonumber\\
& &\times[s^2_{1,||}(s,x)
+\cos(st^*)s_{1,\perp}^2(s,x)],
\end{eqnarray}
where the partition function
$Z_N=\int_0^{N}2s{\cal D}_{N}(s)e^{\alpha s^2}ds$ and
${\cal D}_N(s)$ is the $N$-spin density of states. ${\cal D}_N(x)$ is 
piecewise continuous,
\begin{eqnarray}
{\cal
D}_N(x)&=&\Theta(x)\sum_{p=0}^{E[(N-1)/2]}\Theta(N-2p-x)\times\nonumber\\
& &\times\Theta(x-N+2p+2)d_{N-2p}(x),\\
d_{N-2p}(x)&=&\sum_{k=0}^{p}{{(-1)^k(N-2k-x)^{N-2}}\over{2^{N-1}(N-2)!}}\pmatrix{N\cr
k},
\end{eqnarray}
where $\Theta(x)$ is the Heaviside step function, $\bigl({{n}\atop{m}}\bigr)$
are  binomial distribution coefficients,  and $E(x)$ is the 
largest integer in $x$.
The near-neighbor correlation function ${\cal C}_{12}(t)=\langle {\bf
s}_1(t)\cdot{\bf s}_2(0)\rangle$ is simply obtained from ${\cal C}_{11}(t)$
and the sum rule $\langle s^2\rangle/N=1={\cal
C}_{11}(t)+(N-1){\cal C}_{12}(t)$. 
 
It is useful to write $\delta{\cal C}_{11}(t)={\cal
C}_{11}(t)-\lim_{t\rightarrow\infty}{\cal C}_{11}(t)$.  Since ${\cal
C}_{11}(0)=1$, it suffices to obtain $\delta{\cal C}_{11}(t)$ for
all $t$.  After some algebra, we have reduced 
$\delta{\cal C}_{11}(t)$ to quadrature,
\begin{eqnarray}
\delta{\cal
C}_{11}(t)&=&\!\!\sum_{p=0}^{E[(N-1)/2]}\int_{N-2p-2}^{N-2p}
\!\! ds\cos(st^*)f_{N-2p}(s),\label{C11}\>\>\\
f_{N-2p}(s)&=&\Theta(s){{e^{\alpha
s^2}}\over{Z_N}}\sum_{k=0}^{p}(N-2k-s)^{N-1}g_{k}(s),\label{fN}\\
g_0(s)&=&{{(N-1)a(s)}\over{2^{N-3}s(N+2)!}},\label{g0}\\
a(s)&=&N(s+1)^3-N^2(s+1)+N+s^3-2s,\\
g_k(s)&=&{{(-1)^{k}b_{N-2k}(s)}\over{s2^{N-3}N(N+2)!}}\pmatrix{N\cr k},
\>\>k\ne 0,\\
b_m(s)&=&s^2N(N-1)[3m+s(N+1)]\nonumber\\
& &+[s(N-1)+m]\times\nonumber\\
& &\times[6m^2-N(N+1)(N+2)].\label{endexact}
\end{eqnarray}

To  obtain the long-time asymptotic behavior for  $|\alpha|\ll1$, we 
integrate Eq. (\ref{C11}) 
by parts $N-1$ times.  
We find
\begin{eqnarray}
\delta{\cal
C}_{11}(t)&{{\sim}\atop{t*\gg1}}&\sum_{p=0}^{E(N/2)}h_{p}
\cos[(N-2p)(t^*-\pi/2)]\times\nonumber\\
& &\times (t^{*})^{-N}e^{(N-2p)^2\alpha}/Z_N,\label{longtime}\\
h_p&=&{{[(N-2p)^2-N]}\over{2^{N-3}N}}\pmatrix{N\cr p}\times\nonumber\\
& &\times [1+(p-1)\delta_{p,N/2}].
\end{eqnarray}
 For $N=M^2$, one of the terms in Eq. (\ref{longtime})
vanishes, as for the four-spin ring. \cite{KL}  More
important, the leading long-time behavior $\propto 1/t^{*N}$, which falls off
very rapidly for large $N$. Although Eq. (\ref{longtime}) fails for
$|\alpha|\gg1$,  its behavior for $\alpha\ne0$ is useful to 
illuminate the dramatic difference
between the long-time asymptotic behavior for the  FM and
AFM cases. For the FM case, $\alpha>0$, the dominant behavior
$\propto\cos[N(t^*-\pi/2)]$, as all of the spins oscillate together.  
In the AFM case, $\alpha<0$, the dominant long-time asymptotic behavior is 
given by
the smallest possible oscillation frequency.   As shown in the following,  
${\cal C}_{11}(t)$
is always finite as $T\rightarrow0$.
 
As $T,N\rightarrow\infty$, Liu and M{\"u}ller obtained an analytic form for
${\cal C}_{11}$, but their procedure was not extendable to
finite $N$. \cite{LM}  From the integral representation,
${\cal D}_N(x)=\int_0^{\infty}2p^{1-N}\sin^Np\sin(px)dp/\pi$, we expand $\sin^Np/p^N$ as
$\exp[-N(p^2/6+p^4/180+p^6/2835+\ldots)]$ to  obtain a $1/N$ asymptotic
expansion for ${\cal C}_{11}(t)$ as
$T\rightarrow\infty$, 
\begin{eqnarray}
\lim_{{T\rightarrow\infty}\atop{N\gg1}}{\cal C}_{11}(t)&\sim 
&[1+2e^{-\tilde{t}^2}
(1-2\tilde{t}^2)]/3\nonumber\\
& &+
2[1-e^{-\tilde{t}^2}(1+\tilde{t}^2-\tilde{t}^4/3-2
\tilde{t}^6)]/(5N)\nonumber\\
& &+{{12}\over{175N^2}}\Bigl[1-e^{-\tilde{t}^2}\Bigl(1+\tilde{t}^2-
{{157}\over{36}}\tilde{t}^4+\nonumber\\
& &\>\>{{50}\over{27}}\tilde{t}^6-{{97}\over{108}}\tilde{t}^8+
{7\over{18}}\tilde{t}^{10}\Bigr)\Bigr],\label{coft}
\end{eqnarray} 
where $\tilde{t}=Jt[(N-1)/6]^{1/2}$.
From Eq. (\ref{coft}),  ${\cal
C}_{11}(0)=1$ through order $1/N^2$, as required, and 
${\cal C}_{11}(t)$ has the correct $N\rightarrow\infty$ limit.  \cite{LM} 
However, Eq. (\ref{coft}) is highly accurate for intermediate $N$
values as well.  In Fig. 1, we have plotted $\lim_{T\rightarrow\infty}{\cal
C}_{11}(t)$ for $N=5, 6, 7, 8, 9, 25, 50, 100, \infty$ as a function of
 $Jt[(N-1)/6]^{1/2}$.  The
curves for $N=25, 50, 100, \infty$ were obtained from  Eq. (\ref{coft}),  
whereas the
other curves were generated from the exact solution,
Eqs. (\ref{C11})-(\ref{endexact}). Equation (\ref{coft})
is very accurate at short and long times, even for $N=5$. We note
that for  $\tilde{t}\alt1$, $\lim_{T\rightarrow\infty}{\cal C}_{11}(t)$ is a
universal function of $Jt\sqrt{N-1}$, as suggested by the $1/N$
expansion. 
  Although
the $N\rightarrow\infty$ curve is exact, and the curves for $N=50, 100$ are quite accurate, the curve for $N=25$ exhibits a
small ($\approx1\%$) inaccuracy for $1.2\alt\tilde{t}\alt2.2$. 

\begin{figure}[b]
\includegraphics[width=0.45\textwidth]{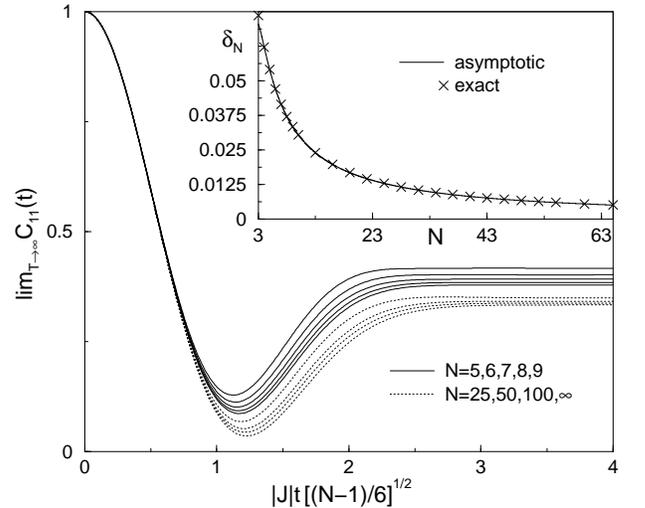}
\caption{Plots of $\lim_{T\rightarrow\infty}{\cal C}_{11}(t)$ versus
$|J|t[(N-1)/6]^{1/2}$, for $N=5,6,7,8,9$ from Eqs. (4)-(9) (solid), and  for
$N=25, 50, 100,\infty$ from Eq. (12) (dashed).  Inset:  Plot of the exact
($\times$) and asymptotic values, Eq. (15), of
$\delta_N$.} 
\end{figure}

At arbitrary $N,t$ and $N\gg1, t\rightarrow\infty$, respectively,
\begin{eqnarray}
\lim_{T\rightarrow\infty}{\cal C}_{11}(t)&=&1/N+(N-1)[\delta_N+f_N(t)],\label{rings}\\
\lim_{t,T\rightarrow\infty}{\cal C}_{11}(t)&{{\sim}\atop{N\gg1}}
&{1\over{3}}+{2\over{5N}}+{{12}\over{175N^2}},\label{tinfinite}\\
\delta_N&{{\sim}\atop{N\gg1}} & {{175N^2-315N+36}\over{525N^2(N-1)}}.\label{deltaN}
\end{eqnarray}
In the inset of Fig. 1, we have compared  the exact values of $\delta_N$ with
those obtained from Eq. (\ref{deltaN}) for $3\le N\le 65$.  For $N=3$, the
differences are noticeable on this curve.  As $N$ increases from 5 to 65, the
relative difference decreases rapidly from
 0.14\% to  0.2 ppm. Since the exact formula for $\delta_N$ with $N$
odd (even) contains  non-trivial contributions of
logarithms of all prime numbers from 3 to $N$  (2 to $N/2$), respectively, the
extremely good numerical agreement between the exact $\delta_N$ and Eq. (\ref{deltaN}) is remarkable. 

In Fig. 2, we plot the
long-time asymptotic limits $\lim_{t\rightarrow\infty}{\cal C}_{11}(t)$ for both the FM and AFM
cases, for $N=5,7,10$, as functions of $|\alpha|$.  The  FM values
approach 1 as $T\rightarrow0$.  But, the
low-$T$ AFM values rapidly approach their $T=0$ asymptotic limit, 
${1\over{3}}$.
This indicates the spins are strongly  frustrated, since
one might naively expect the AFM $\lim_{t\rightarrow\infty}{\cal C}_{11}(t)=0$, as for the four-spin ring. \cite{KL}  In that case,
the unfrustrated alternating spin direction configuration was possible.  In this AFM
case, however, the $N\ge3$ spins are as frustrated as $T\rightarrow0$ as
are an infinite number of them  as $T\rightarrow\infty$.

\begin{figure}[floatfix]
\includegraphics[width=0.45\textwidth]{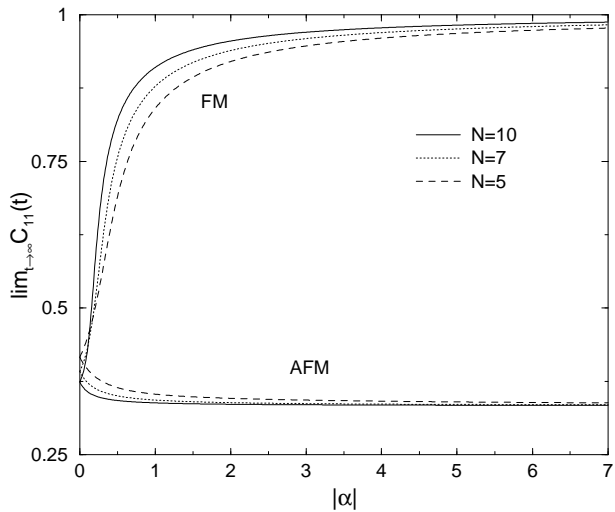}
\caption{Plots of $\lim_{t\rightarrow\infty}{\cal C}_{11}(t)$ versus $|\alpha|$ for
$N=5,7,10$ and for both ferromagnetic and antiferromagnetic couplings.}
\end{figure}

We now examine the low-$T$ AFM dynamics.  In this case, the
dominant contibutions to the numerator of ${\cal C}_{11}$ and to $Z_N$  arise
from $s\ll1$.  The leading contributions to these integrals are
$P_Ns^2e^{-|\alpha|s^2}$ and $Q_Ns^2e^{-|\alpha|s^2}$, respectively, where
$P_N$ and $Q_N$ are complicated functions of $N$, but for arbitrary $N\ge3$, 
$P_N/Q_N={2\over{3}}$.
 Thus, as
$T\rightarrow0$,  the Fourier transform $\delta{\cal C}_{11}(\omega)$ of $\delta{\cal
C}_{11}(t)$ attains the exact uniform asymptotic form,
\begin{eqnarray}
|\alpha|^{-1/2}\delta{\cal
C}_{11}(\omega)&{{\sim}\atop{T\rightarrow0}}&{{8}\over{3\sqrt{\pi}}}\tilde{\omega}^2e^{-\tilde{\omega}^2},\label{exactAFM}
\end{eqnarray}
where $\tilde{\omega}=|\alpha|^{-1/2}\omega/|J|$.
  For $N\ge5$,
corrections to Eq. (\ref{exactAFM}) are of ${\cal O}(|\alpha|^{-1})$, whereas for $N=4$,
they are of  ${\cal O}(|\alpha|^{-1/2})$.  
In Fig. 3, we show the
frequency dependence of the low-$T$ AFM mode.  Since
${\cal C}_{11}(t)$ for fixed $N$ is a uniform function of $|J|t/|\alpha|^{1/2}$,
the amplitude of $\delta{\cal C}_{11}(\omega)$ must also be scaled by $|\alpha|^{-1/2}$ for the curves to nearly
coincide.  Here we plot the curves at $\alpha=-5,-10$ for $N=4,7,10$, obtained
from our exact solution, Eqs. (\ref{C11}) - (\ref{endexact}). The exact asymptotic form,
Eq. (\ref{exactAFM}) is also plotted for comparison.
 The
$N=4$ curves nearly coincide with the others, but do show some small
dependence upon $T$.  However, the $N=7$ and 10 curves are nearly
identical and  independent of $T$, and are nearly indistinguishable from the
asymptotic curve for $|\alpha|=10$,  when scaled in this manner. 

\begin{figure}[floatfix]
\includegraphics[width=0.45\textwidth]{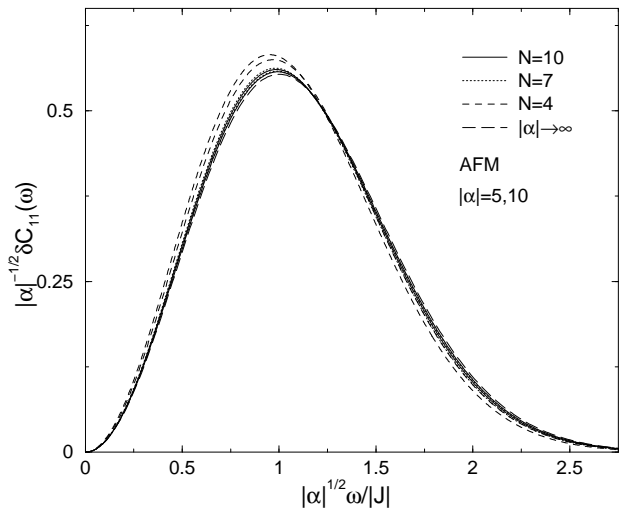}
\caption{Plots of the low-$T$ $|\alpha|^{-1/2}\delta{\cal C}_{11}(\omega)$ versus
$|\alpha|^{1/2}\omega/|J|$ for the antiferromagnetic cases with $N=4,7,10$ at
$\alpha=-5,-10$.  Also shown is the exact $T\rightarrow0$ limit, Eq. (16).}
\end{figure}

Inverting the Fourier transform, the low-$T$ AFM ${\cal C}_{11}(t)$ asymptotically
approaches 
\begin{eqnarray}
{\cal
C}_{11}(t)&{{\sim}\atop{T\rightarrow0}}&[1+2e^{-\overline{t}^2}(1-2\overline{t}^2)]/3,
\end{eqnarray} 
where $\overline{t}=t^*/(2|\alpha|^{1/2})$.
Hence the AFM $\lim_{T\rightarrow0}{\cal C}_{11}(t)$ is a uniform function of $(JT)^{1/2}t$, independent of $N$.
Moreover, it has precisely the same form as does
$\lim_{N,T\rightarrow\infty}{\cal C}_{11}(t)$, Eq. (\ref{coft}), pictured
in Fig. 1, except for the different scaling factor.

We now turn to the FM case as $T\rightarrow0$.  From Eq. (\ref{longtime}),
we expect that all of the spins will be oscillating together with frequency
$NJ$.  But, this is strictly true only in the $T\rightarrow0$ limit.  At finite
$\alpha$, the oscillation frequency  deviates from this value slightly.
To determine more precisely the actual nature of the mode, we note that for
$\alpha>1$, the peaks in the integrand and in $Z_N$ both occur
within $N-2\le s\le N$.  As $T\rightarrow0$, we evaluate $Z_N$ by integration
by parts, leading to the exact low-$T$ limit of the FM Fourier transform, 
\begin{eqnarray}
\delta{\cal
C}_{11}(\omega)&{{\sim}\atop{T\rightarrow0}}&A_Nf_N(\omega^*)/f_N(\omega^*_N),\label{FMexact}
\end{eqnarray}
where $A_N=2e[(N-1)/e]^N/N!$,
$\omega^*=\omega/J$, and $\omega_N^*$,  the position of the exact maximum in
$f_N(\omega^*)$, is an eigenvalue of
 $2\alpha s-(N-2)/(N-s)+d\ln[g_0(s)]/ds=0$.  $f_N$ and $g_0$ are given by
Eqs. (\ref{fN}) and (\ref{g0}), respectively.  Approximately, 
$\omega^*_{N}\approx N -(N-1)/(2N\alpha)$, and $f_N(\omega^*)$ has a width characterized by a normal Gaussian
distribution parameter
$\sigma_N=\sqrt{N-1}/(2N\alpha)$.  
Thus,
\begin{eqnarray}
\delta{\cal
C}_{11}(\omega)&{{\approx}\atop{\alpha\gg1}}&A_N
\exp[-(\omega^*-\omega_{N}^{*})^2/(2\sigma^2_{N})].\label{FMapprox}
\end{eqnarray}
 In
 this Gaussian approximation, 
  $\omega^*_{N}$, $\sigma_N$, and $A_N$ are correctly given to their
 respective leading orders in $1/\alpha$, but the skewness of the 
peak is not accurately 
represented. 
In Fig. 4, we have therefore plotted the scaled low-$T$  $\delta{\cal C}_{11}(\omega)/A_N$ versus
$(\omega^*-\omega^*_{N})N\alpha/\sqrt{N-1}$, for $N=8$ at $\alpha=5,10$, and
 for $N=5$ at $\alpha=5, 10, 28$,  using our exact
formulas, Eqs. (\ref{C11}) - (\ref{endexact}). The $N=5$ curves nearly
 coincide, as do the $N=8$ curves, and the scaled height approaches 1 as
 $T\rightarrow0$.  As mentioned
 previously, the asymmetry of the curves decreases with increasing $N$, a
 feature not incorporated in the Gaussian approximation.  Neverthess this
 scaling procedure
 works remarkably well, and  $N=5$
is close to the large $N$ limit, even for the FM case. 

\begin{figure}[floatfix]
\includegraphics[width=0.45\textwidth]{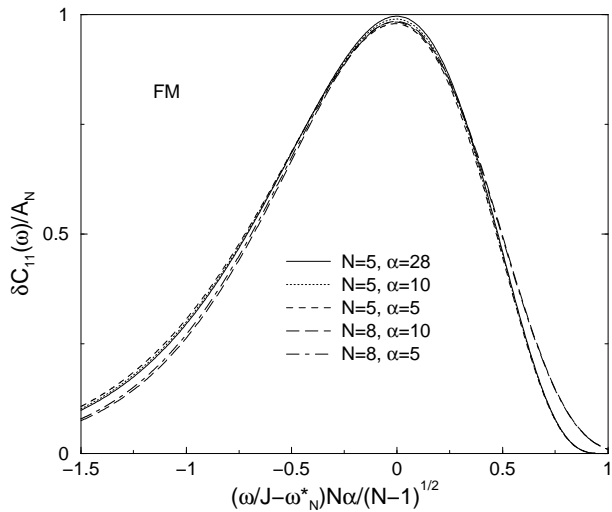}
\caption{Plots of the low-$T$ $\delta{\cal C}_{11}(\omega)/A_N$ versus
$(\omega/J-\omega^*_N)N\alpha/\sqrt{N-1}$, where
$\omega^*_N=N-(N-1)/(2N\alpha)$ and  $A_N=2e[(N-1)/e]^N/N!$, for the FM cases with $N=8$,
$\alpha=5,10$, and $N=5$, $\alpha=5, 10, 28$.}
\end{figure}

 Inverting the low-$T$ FM Fourier transform, Eq. (\ref{FMapprox}),
\begin{eqnarray}
{\cal
C}_{11}(t)&{{\approx}\atop{T\rightarrow0}}&1-{{B_N}\over{\alpha}}
\Bigl[1-e^{-t^{*2}\sigma_N^2/2}\cos(\omega^*_{N}t^*)\Bigr],\label{FMlowT}
\end{eqnarray}
 where $B_N=A_N\sqrt{\pi(N-1)/2}/N$.
   As
$N\rightarrow\infty$, $B_N\rightarrow 1/N$.  The deviation of the long-time
   limit of Eq. (\ref{FMlowT})
agrees to within 1\% with our results presented in Fig. 2.  The decay in time of the
 mode in this approximation is  Gaussian,
with a lifetime $\tau_N=1/\sigma_N$.  Thus, as $T\rightarrow0$,  the 
oscillations are characterized by a frequency that approaches  $NJ$ linearly
   in $T$, a lifetime that diverges as $1/T$, and an amplitude that vanishes as $T$. 

Finally, we remark that our exact results  for  $\lim_{T\rightarrow\infty}{\cal
C}_{11}(t)$ in the $N$-spin equivalent neighbor model may also lead to the correct $T\rightarrow\infty$ limit
of the near-neighbor correlation function ${\cal
C}_{12}(t)$ of $N$-spin rings with classical Heisenberg near-neighbor exchange
coupling.  From Eq. (\ref{rings}) and the sum rule,  we
conjecture that, as for the $N$-spin equivalent neighbor model,   
\begin{equation}
\lim_{T\rightarrow\infty}{\cal C}_{12}(t)=1/N-\delta_N-f_N(t)
\label{conjecture}  
\end{equation}
for the $N$-spin ring. In support of this conjecture,
we have studied the $N$-spin systems consisting of $M$ spins ${\bf s}_j'$
coupled to $N-M$ spins ${\bf s}_i$ on a ring.   Letting 
${\bf s}=\sum_{i=1}^{N-M}{\bf
s}_i$ and ${\bf s}'=\sum_{j=1}^{M}{\bf s}_j'$, the Hamiltonian is
\begin{equation}
H=-J\sum_{i=1}^{N-M}{\bf s}_i\cdot{\bf s}_{i+1}-J'{\bf s}\cdot{\bf s}',
\end{equation}
where ${\bf s}_{N-M+1}={\bf s}_1$, which is integrable. \cite{SKMWT}  Setting ${\cal C}'(t)=\langle 
{\bf s}'(t)\cdot{\bf s}'(0)\rangle$,  we found for $N-M=3$, $M=1, 2, 3$, and
for $N-M=4$, $M=1, 2$, that
\begin{eqnarray}
\lim_{T\rightarrow\infty}{\cal C}'(t)&=&M^2/N+M(N-M)[\delta_N+f_N(t)].
\end{eqnarray}
Most significant of these is the result for the four-spin ring with $M=2$.  We
also showed previously that Eq. (\ref{conjecture}) is valid for four- and three-spin rings. \cite{KL,AK}

In summary, we have solved exactly for the time autocorrelation function
${\cal C}_{11}(t)$ of a
spin in the $N$-spin equivalent neighbor model.  As $T\rightarrow\infty$,
${\cal C}_{11}(t)\propto t^{-N}$ for $Jt\gg1$,  curves for different $N$ exhibit very
similar shapes when plotted as functions of $|J|t[(N-1)]^{1/2}$, and an
accurate $1/N$ asymptotic expansion  is found. At
low $T$,  the antiferromagnetic ${\cal C}_{11}(t)$ is a universal function of $tT^{1/2}$,
independent of $N$, and exhibiting strong frustration. For the
ferromagnetic case at low $T$, there is a single mode with a peak position at
$\approx NJ-k_BT(N-1)/N$, and a shape  that fits  an accurate scaling
relation.  We conjecture that the $T\rightarrow\infty$ limit of the 
near-neighbor correlation functions ${\cal C}_{12}(t)$
 for the $N$-spin equivalent neighbor model and for the  $N$-spin ring 
may be identical.

\end{document}